**Electronic bandgap and exciton binding energy of layered semiconductor TiS$_3$**

*Aday J. Molina-Mendoza, Mariam Barawi, Robert Biele, Eduardo Flores, José R. Ares, Carlos Sánchez, Gabino Rubio-Bollinger, Nicolás Agraït, Roberto D'Agosta\*, Isabel J. Ferrer\*, and Andres Castellanos-Gomez\**

A. J. Molina-Mendoza[1], Prof. G. Rubio-Bollinger[1,2,3], Prof. N. Agraït[1,2,3,4],
[1]Dpto. de Física de la Materia Condensada, Universidad Autónoma de Madrid, Campus de Cantoblanco, E-28049 Madrid, Spain
[2]Instituto de Ciencia de Materiales "Nicolás Cabrera", Campus de Cantoblanco, E-18049 Madrid, Spain
[3]Condensed Matter Physics Center (IFIMAC), Universidad Autónoma de Madrid, Campus de Cantoblanco, E-28049 Madrid, Spain
[4]Instituto Madrileño de Estudios Avanzados en Nanociencia (IMDEA-nanociencia), Campus de Cantoblanco, E-18049 Madrid, Spain

M. Barawi[5], E. Flores[5], Prof. J. R. Ares[5], Prof. C. Sánchez[2,5], Prof. I. J. Ferrer[2,5]
[2]Instituto de Ciencia de Materiales "Nicolás Cabrera", Campus de Cantoblanco, E-18049 Madrid, Spain
[5]Materials of Interest in Renewable Energies Group (MIRE Group), Dpto. de Física de Materiales, Universidad Autónoma de Madrid, Campus de Cantoblanco, E-28049 Madrid, Spain
E-mail: isabel.j.ferrer@uam.es

R. Biele[6], Prof. R. D'Agosta[6,7]
[6]ETSF Scientific Development Center, Dpto. de Física de Materiales, Universidad del País Vasco, E-20018 San Sebastián, Spain
[7]IKERBASQUE, Basque Foundation for Science, E-48013, Bilbao, Spain
E-mail: roberto.dagosta@ehu.es

Dr. A. Castellanos-Gomez[4]
[4]Instituto Madrileño de Estudios Avanzados en Nanociencia (IMDEA-nanociencia), Campus de Cantoblanco, E-18049 Madrid, Spain
E-mail: andres.castellanos@imdea.org

We present a study of the electronic and optical bandgap in layered TiS$_3$, an almost unexplored semiconductor that has attracted recent attention because of its large carrier mobility and inplane anisotropic properties, to determine its exciton binding energy. We combine scanning tunneling spectroscopy and photoelectrochemical measurements with random phase approximation and Bethe-Salpeter equation calculations to obtain the electronic and optical bandgaps and thus the exciton binding energy. We find experimental values for the electronic bandgap, optical bandgap and exciton binding energy of 1.2 eV, 1.07 eV and 130 meV, respectively, and 1.15 eV, 1.05 eV and 100 meV for the corresponding theoretical results. The exciton binding energy is orders of magnitude larger than that of common semiconductors and comparable to bulk transition metal dichalcogenides, making $TiS_3$ ribbons a highly interesting material for optoelectronic applications and for studying excitonic phenomena even at room temperature.

**1. Introduction**

In the last four years there has been a revival of the interest on layered semiconductor materials. For example, Mo- and W- based dichalcogenides (transition metal dichalcogenides, TMDCs hereafter), have been used in field-effect transistors, optical devices, logic circuits or memories.[1-12] Layered Mo- and W- TMDCs are very unusual semiconductors due to their large exciton binding energy (ranging from 70 meV in bulk up to 900 meV in monolayer) compared to other widely used semiconductors like silicon, germanium, gallium arsenide or groups II-VI and III-V (ranging from 1 meV up to 60 meV).[13-25] Narrow and pronounced photoluminescence emission even at room temperature can be observed due to this unusually large exciton binding energy, making it possible to study interesting optical phenomena such as charged excitons or valley polarization.[26-28]

Mo- and W-based dichalcogenides, however, have very poor optical absorption for wavelengths larger than 950 nm, hampering its use in applications that require a wider range of absorption (night vision imaging systems, thermography or communications). A direct bandgap is also a desirable characteristic in terms of optical devices applications due to a more efficient light absorption, feature that is only present in the TMDCs at the monolayer limit. Therefore, new materials solving these issues are highly attractive. TiS$_3$, a layered semiconducting trichalcogenide with a direct bulk optical bandgap of 1 eV, has recently gained attention due to its interesting properties like low cost production, abundant and nontoxic elements and easy synthesis process, as well as due to potential applications in different fields like solar cells, photoelectrochemical cells, cathodes in batteries, hydrogen storage, thermoelectric conversion devices and optoelectronics.[29-40] TiS$_3$ grows in layered ribbon-like shape, showing n-type conduction behavior with current on/off ratios of $10^4$, mobilities up to 73 cm$^2$ V$^{-1}$ s$^{-1}$ and it has already been used in field-effect transistors, showing ultrahigh photoresponse up to 2910 A W$^{-1}$. Although its optical properties have been recently studied, little is still known about its electronic band structure, and thus its electronic bandgap and exciton binding energy.

Here we study the electronic and optical bandgap in bulk TiS$_3$ ribbons (30 to 130 layers thick). We measure the electronic bandgap ($E_{g,el}^{\exp}$) by means of scanning tunneling spectroscopy (STS) at room temperature and the optical bandgap ($E_{g,opt}^{\exp}$) by photoelectrochemical measurements, obtaining $E_{g,el}^{\exp} = 1.20 \pm 0.08 eV$ and $E_{g,opt}^{\exp} = 1.07 \pm 0.01 eV$, respectively. These values are in good agreement with the results of random phase approximation (RPA) and Bethe-Salpeter equation (BSE) calculations, which give $E_{el}^{th} = 1.15 eV$ for the electronic bandgap and $E_{opt}^{th} = 1.05 eV$ for the optical one.

From these two values we are able to estimate the exciton binding energy of TiS$_3$ ribbons as the difference between the electronic and the optical bandgap $E_{exc}^{exp} = E_{g,el}^{exp} - E_{g,opt}^{exp} = 130\,meV$, also in good agreement with the one obtained from the calculations $E_{exc}^{th} = 100\,meV$. This binding energy is 1-2 orders of magnitude larger than that of common semiconductors (silicon, germanium, gallium arsenide…) and comparable to bulk TMDCs, making TiS$_3$ a highly interesting and promising material for studying interesting excitonic physical phenomena like the ones recently found in layered TMDCs.

## 2. Results and Discussion

We experimentally investigated the electronic and optical bandgap of TiS$_3$ ribbons by STS and photoelectrochemical measurements. The samples employed in the STS study are prepared by transferring TiS$_3$ ribbons onto a flame-annealed gold substrate. This process is illustrated in **Figure 1a** and explained in detail in the experimental section. Briefly, a viscoelastic stamp is gently brought into contact with the TiS$_3$ ribbons, grown on the surface of a Ti disc by sulphuration at 500 ºC. By suddenly peeling the stamp off, several TiS$_3$ ribbons are transferred to the stamp, making it possible to deposit the ribbons onto the gold substrate by gently pressing the stamp against the gold surface and slowly peeling it off. Exfoliation occurs trough the (001) plane, contained by the a and b axis of the lattice (see Figure S5 of the Supporting information for schematic drawing of the TiS$_3$ crystal structure). Optical microscopy and atomic force microscopy (AFM) images of the sample are shown in Figure 1b. The transferred ribbons are typically 25 nm to 105 nm thick (30 to 130 layers), with an average thickness of 60 nm. The surface area covered by TiS$_3$ is around 5% of the total sample area.

The fabricated sample is studied with a homebuilt scanning tunneling microscope (STM) operated at room temperature and ambient conditions.[41] The regions of the sample uncovered with TiS$_3$ are easily distinguished by their high differential conductance at zero bias in the current vs. voltage curves (called IVs hereafter) as well as by their corrugated topography (both features characteristic of evaporated gold films). When the STM tip is placed onto a TiS$_3$ ribbon the tunneling IVs show a wide region with zero differential conductance, as expected for a semiconductor material.[42, 43] Although we found it difficult to scan on the surface of the TiS$_3$ (probably due to poor TiS$_3$-Au adhesion), the topography images show flat terraces with sharp steps, as expected for a layered material as TiS$_3$ (see inset with blue frame in Figure 1c). Since these features are not unique to TiS$_3$ ribbons, i.e., Au substrate could also present terraces and steps in its (111) reconstruction (although the employed substrate is simply evaporated gold, not reconstructed), we further confirm the presence of TiS$_3$ by measuring IVs (see the Experimental Section for details) as for TiS$_3$ the IVs show a clear drop of the current down to zero when the bias voltage shifts the Fermi level within the TiS$_3$ bandgap, while for gold the curve profile is that expected for a metal-vacuum-metal junction. Figure 1c compares STS IVs acquired on TiS$_3$ ribbons and on gold.

The measured STS spectra strongly depends on the exact arrangement of the atoms involved in the tunneling process, meaning that at ambient conditions the atomic diffusion and the thermal drift yields fluctuations between different measured IV traces. In order to overcome that issue, we recorded more than 1000 IVs in different TiS$_3$ ribbons at different distances, *i.e.*, different tunneling current setpoints, to run a statistical analysis with a 2D histogram (see the experimental section for details).[44, 45] We have chosen a set of 205 IVs among all the recorded curves to build the 2D histogram illustrated in **Figure 2a** (see Supporting Information for the complete set). A line profile along zero current (dashed line in Figure 2a) is taken to represent the number of counts as a function of tip voltage. This one dimensional histogram (Figure 2b) is used to determine the valence and conduction bands values as the voltage value at which the counts are 50% of the highest count (vertical red lines

This is the post-peer reviewed version of the following article:
A.J. Molina-Mendoza *et al*. "Electronic bandgap and exciton binding energy of layered semiconductor TiS3"
Advanced Electronic Materials vol. 1, issue 9, page 1500126 (2015)
DOI: 10.1002/aelm.201500126
Which has been published in final form at:
http://onlinelibrary.wiley.com/doi/10.1002/aelm.201500126/abstractin Figure 2b). Using this method the measured value for the valence band is $E_{VB} = -0.64 \pm 0.06\,eV$, while for the conduction band the measured value is $E_{CB} = 0.47 \pm 0.06\,eV$. It is important to note that the IVs show a shift of the Fermi level towards the conduction band, meaning that the material is a n-doped semiconductor. The n-type doping of this compound has already been studied with transport measurements and found to be related to the presence of sulphur vacancies, similar to other semiconducting dichalcogenides.[36, 46-49]

Another aspect of STS measurements at ambient conditions that needs to be considered is that the IVs are thermally broadened, which leads to a slightly underestimation of the bandgap value. This band broadening, as indicated by S. Crampin et al. in Ref. [50], is induced by the temperature dependence of the Fermi-Dirac function and it is obtained from the convolution of the differential conductivity of STS with the derivative of the Fermi-Dirac function. As a result of this, STS measurements performed at a finite temperature $T$ yield traces that are broadened by $E_t \approx 3.5 k_B T$ (where $k_B$ is the Boltzmann constant), meaning that the bandgap will be reduced by this magnitude which, in our case, has a value of $E_t = 90\,meV$.[50] Adding the thermal broadening quantity to the measured bandgap value we see that the electronic bandgap of TiS$_3$ ribbons is: $E_{g,el}^{\exp} = E_{CB} - E_{VB} + E_t = 1.20 \pm 0.08\,eV$.

The optical bandgap was determined by photoelectrochemical measurements using a three-electrode cell in an aqueous solution (see experimental section for details) with the setup illustrated in **Figure 3a**. Photocurrent spectral response density of TiS$_3$ at 0.5 V (*vs.* Ag/AgCl) is represented in a Tauc plot (Figure 3b).[51] The optical bandgap is determined by the point where a linear fit of the data crosses the x-axis, *i.e.*, the wavelength or photon energy. This method also provides information about the nature of the gap, direct or indirect, depending on the shape of the spectra. In these plots the dependence of an optical magnitude *A* (absorption coefficient, photocurrent, ...) on the photon energy (or wavelength) defines the type of the optical energy transition between bands. The linear fit of $(A \cdot h\upsilon)^n$ *vs.* $h\upsilon$ for n = 2 hints direct allowed transition and for n = 1/2 implies indirect allowed transition. For TiS$_3$ two direct transitions can be observed corresponding to $1.07 \pm 0.01 eV$ and $1.29 \pm 0.01 eV$. The transition corresponding to the formation of the first exciton state is the one depicted in Figure 3b: $E_{g,opt}^{\exp} = 1.07 \pm 0.01 eV$. These results are in good agreement with the previously reported experimental values obtained from photocurrent response and optical absorption measurements, obtained for TiS$_3$ samples grown under different conditions and different precursor materials.[30, 31]

Considering the measured values for both the electronic and optical bandgap we can calculate the corresponding exciton binding energy as the difference between the two bandgaps: $E_{exc}^{\exp} = E_{g,el}^{\exp} - E_{g,opt}^{\exp} = 130 meV$.

To better understand and interpret our experimental findings, we performed state-of-the-art density functional theory (DFT) calculations in combination with many-body techniques. Namely, we first investigated the electronic structure of bulk TiS$_3$ within DFT (as the experimentally studied TiS$_3$ ribbons are relatively thick). As DFT tends to underestimate the electronic bandgap, we performed G$_0$W$_0$ calculations in order to get an accurate value for the electronic bandgap.

To access the optical gap, one has to include the interaction between the excited electron in the conduction band with the hole created in the valence band. For this we solved the Bethe-Salpeter equation (BSE) starting from the GW corrected DFT results.[52, 53] In **Figure 4** the absorption spectra of bulk TiS$_3$ are shown with (BSE) and without (RPA) electron-hole interaction. The absorption peak in the RPA spectrum at $E_{el}^{th} \cong 1.15$ eV coincides with the electronic bandgap, while the peak of the BSE spectrum at $E_{op}^{th} \cong 1.05$ eV gives the optical gap. The peaks at ~1.2 eV correspond to an optical transition to a higher energy level in the conduction band, as mentioned in Ref.[36]. Hence, the exciton binding energy is $E_{exc}^{th} \cong 100$ meV, in excellent agreement with the experimental results.

## 3. Conclusion

In summary, we have measured both the electronic and optical bandgap of layered TiS$_3$ ribbons by combining STS at room temperature and photoelectrochemical measurements, finding values of 1.2 eV for the electronic bandgap and 1.07 eV for the optical bandgap. We also estimate the exciton binding energy of TiS$_3$ ribbons as the difference between the two measured bandgaps, obtaining a value of 130 meV. These experimental results are in good agreement with the ones obtained by RPA and BSE calculations, 1.15 eV for the electronic bandgap, 1.05 eV for the optical bandgap and 100 meV for the exciton binding energy. The exciton binding energy of TiS$_3$ ribbons is 1-2 orders of magnitude larger than that of common semiconductors like silicon, germanium or gallium arsenide, and comparable to bulk TMDCs, making it a prospective material for optical spectroscopy experiments at room temperature.

## 4. Experimental Section

*TiS$_3$ growing:* TiS$_3$ samples were grown by direct sulphuration of Ti discs *(Goodfellow 99,9%, Ø=10mm)* in a vacuum sealed ampoule (10$^{-3}$ mbar) with sulphur powder *(Merk 99,75%)* at 500 ºC for 20 hours. At this temperature, sulphur pressure reaches 2 bars. Ti discs had been previously etched in HF:HNO$_3$ mixture (4% wt:30% wt) to remove the possible impurities accumulated in their surfaces.

*Sample preparation:* The samples used for STS measurements (exfoliated TiS$_3$ ribbons) are prepared by placing a dry viscoelastic stamp (*Gelfilm from GelPak*)® on bulk TiS$_3$ (grown by sulfuration of Ti discs at 500 ºC) and peeling it off fast.[30, 31] The stamp is placed on a flame-annealed gold substrate afterwards and removed slowly, leaving exfoliated material on the surface (see Figure 1a). The sample is then annealed at 170 ºC in air during 3 hours to remove organic residues present on the surface. Characterization with AFM yields an average thickness for the ribbons of 60 nm, which corresponds to ~75 layers (see Supporting Information). The samples used for optical measurements are those just sulphurated discs which are connected to the potentiostat unit by the Ti substrate. These sulphurated discs are composed of interconnected TiS$_3$ ribbons with an average thickness of ~100 nm – 200 nm.

*Scanning tunneling spectroscopy measurements:* STS measurements for determining the electronic bandgap were carried out with a home-built STM with mechanically cut gold tips.[40] A tunneling current setpoint of ~100 pA is set to avoid band-bending effects in the measurements. The probability of finding a TiS$_3$ ribbon with the STM tip is high enough since ~5% of the surface is covered by TiS$_3$ (Figure 1b). In order to find TiS$_3$ ribbons, the STM tip is placed randomly on the sample, running IVs at different areas of the total piezotube scan range. It is easy to distinguish the bare gold region from the TiS$_3$ by *in situ* analyzing the IVs (as can be seen in Figure 1c), the IV profile on gold show high differential conductance around zero bias voltage, while for TiS$_3$ the region with zero conductance is extended between two well defined bias voltage values, corresponding to the valence and conduction bands (see Results and Discussion section for details).

*2D histogram:* The recorded IVs are plot in absolute value in a 2D histogram (Figure 2a), used to determine the valence and conduction bands values. This 2D histogram is built by discretizing the bias voltage and current axes into N number of bins, forming a NxN matrix (400x400 in this case).[43] Each data point with current and voltage values within the interval of one bin adds one count to it, representing each count with a colormap. A histogram with the current bins around zero current (0 pA < I ≤ 0.5 pA) is plotted afterwards (Figure 2b), choosing the bias voltage value corresponding to 50% of the maximum counts as the valence and conduction bands values (see results and discussion for details). Figure S3 and Figure S4 of the Supporting Information show 2D histograms built from synthetic curves that clarify the way in which the 2D histograms are constructed.

*Photoelectrochemical measurements:* the photoelectrochemical measurements for determining the optical bandgap were performed in a three-electrode glass cell with a quartz window in an aqueous solution of 0.5 M $Na_2SO_3$ buffered a pH = 9.5 as illustrated in Figure 3b. $TiS_3$ ribbons were used as working photoanode. The counter electrode was a platinum sheet, and the reference one was an Ag/AgCl electrode. A halogen lamp was used as visible light source, which is coupled to a Bausch and Lomb monochromator. Photocurrents at 0.5 V (Ag/AgCl) under different photon energies have been measured by using a potentiostat-galvanostat PGSTAT302N to obtain the photocurrent spectral response. During the experiment, an argon flow of 20 sscm was passed through the top of the cell.

*DFT calculations are combined with many-body techniques*: Both for Ti and S, the exchange-correlation potential is described self-consistently within the generalized gradient approximation throughout the Perdew-Burke-Ernzerhof's functional (PBE). For S the Martins-Troulliers' pseudopotential is used, while for Ti the Goedecker-Hartwigsen-Hutter-Teter's pseudopotential, including semi-core states for the valence electrons, is used.[54, 55] These pseudopotentials are norm-conserving and scalar relativistic. By starting from the lattice parameters provided in Ref. [36] we have optimised the atomic positions with a residual force after relaxation of 0.001 a.u. using the Broyden–Fletcher–GoldfarbShann's procedure. The kinetic energy cutoff for the plane wave basis set is put at 180 Ry, while the cutoff for the charge density is 720 Ry. The sampling of the Brillouin zone is $10 \times 10 \times 10$ according to the Monkhorst−Pack scheme. First-principles electronic structure calculation and structure optimisation have been performed by the DFT pseudopotentials plane-wave method as implemented in the PWSCF code of the Quantum-ESPRESSO package.[56]

To go beyond the DFT-PBE level the electronic structure is further corrected by means of non-self-consistent $G_0W_0$ approximation. The screening in the $G_0W_0$ calculation is treated within the so-called plasmon pole approximation.[57, 58] The local field effects in the screening calculations have been taken into account and we carefully converged the electronic quasiparticle gap.

When looking at two-particle properties such as absorption of light, one has to include the interaction between the excited electron in the conduction bands with the hole created in the valence band. For this we solved the BSE using the recursive Haydock Method to access the optical gap.[52, 53, 59]

To construct the BS kernel in the static approximation we considered 30 valence and 70 conduction bands.[60] The position of the first peak in the optical spectrum, which corresponds to the optical gap, has been carefully converged for example with respect to the *k*-point sampling, the components to be summed in the exchange part and those of the screened coulomb potential of the BSE kernel. In order to make the first peak in the BSE spectrum better visible, in Figure 4 we have plotted only the transitions between the highest occupied and the lowest unoccupied electronic states.

. The plane-wave code Yambo is used to calculate quasiparticle corrections and optical properties with and without electron-hole effects.[61]

**Supporting Information**

Supporting Information is available from the Wiley Online Library or from the author.

**Acknowledgements**

This work was supported by MICCINN/MINECO (Spain) through the programs MAT2011-25046, MAT2014-57915-R, BES-2012-057346, FIS2011-23488 and CONSOLIDER-INGENIO-2010 (CSD-2007-00010), Comunidad de Madrid (Spain) through the programs NANOBIOMAGNET (s2009/MAT-1726) and S2013/MIT-3007 (MAD2D), and the Fundacion BBVA through the fellowship "I Convocatoria de Ayudas Fundacion BBVA a Investigadores, Innovadores y Creadores Culturales" ("Semiconductores ultradelgados: hacia la optoelectronica flexible"). Authors from MIRE Group want to thank to MINECO (Spain) for support through the project MAT2011-22780, CONACyT (México) for the Ph D.Grant provided to E.F. and the technical support from Mr. F. Moreno. R.B. and R.D'A. acknowledge the financial support of the Grupos Consolidados UPV/EHU del Gobierno Vasco (IT579-13), CONSOLIDER-INGENIO-2010: NANOTherm (CSD2010-00044) and useful discussions with Dr. H. Huebener.

Received: ((will be filled in by the editorial staff))
Revised: ((will be filled in by the editorial staff))
Published online: ((will be filled in by the editorial staff))

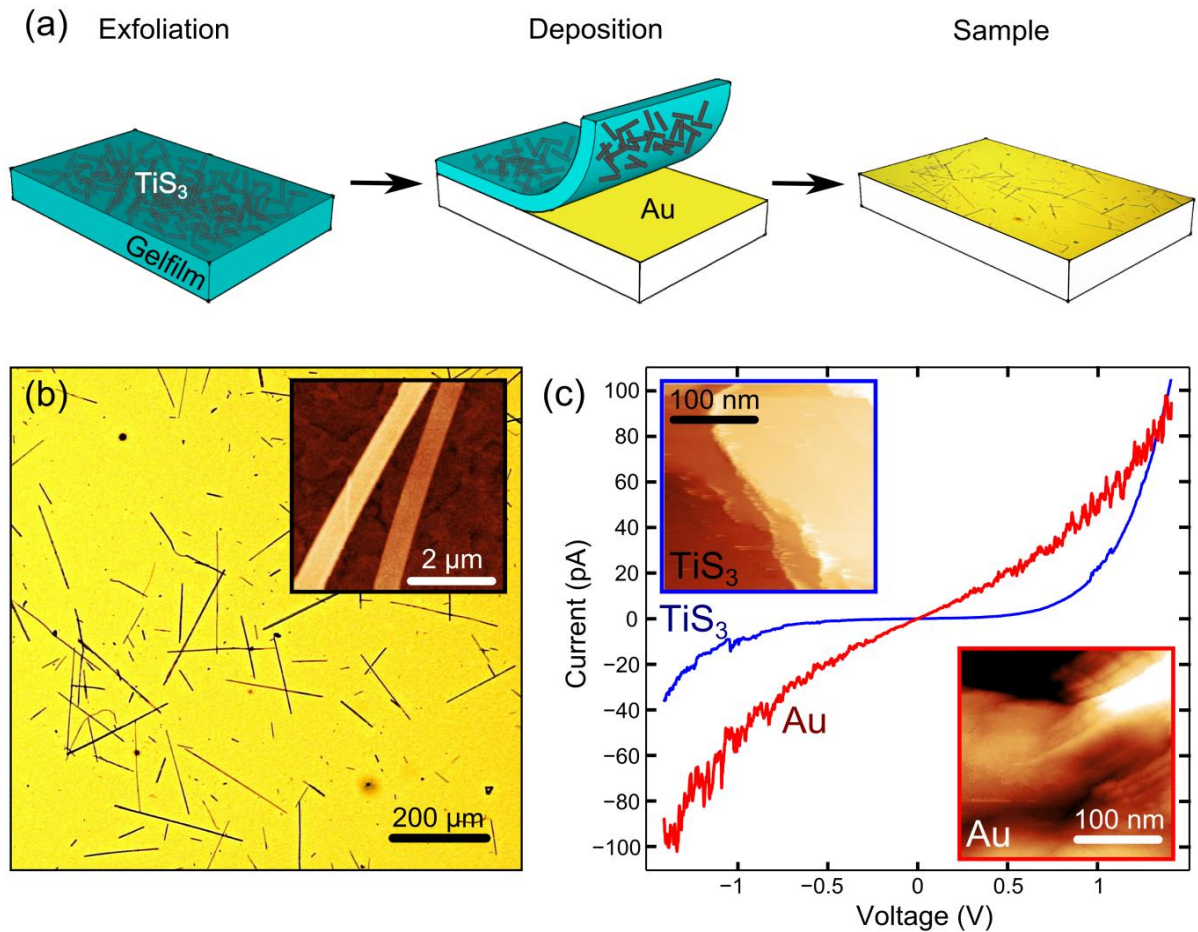

**Figure 1.** (a) Schematic diagram of the sample preparation for STS measurements. A Gelfilm is first placed onto a bulk TiS$_3$ sample, then it is pressed against a gold substrate and peeled off, leaving exfoliated TiS$_3$ ribbons on the surface. (b) Optical microscopy image of TiS$_3$ ribbons on a gold substrate. Inset: AFM topographic image of two TiS$_3$ ribbons. (c) STS current-voltage characteristics of TiS$_3$ (blue) and bare gold (red). Insets: STM topographic images of TiS$_3$ (blue frame) and bare gold (red frame).

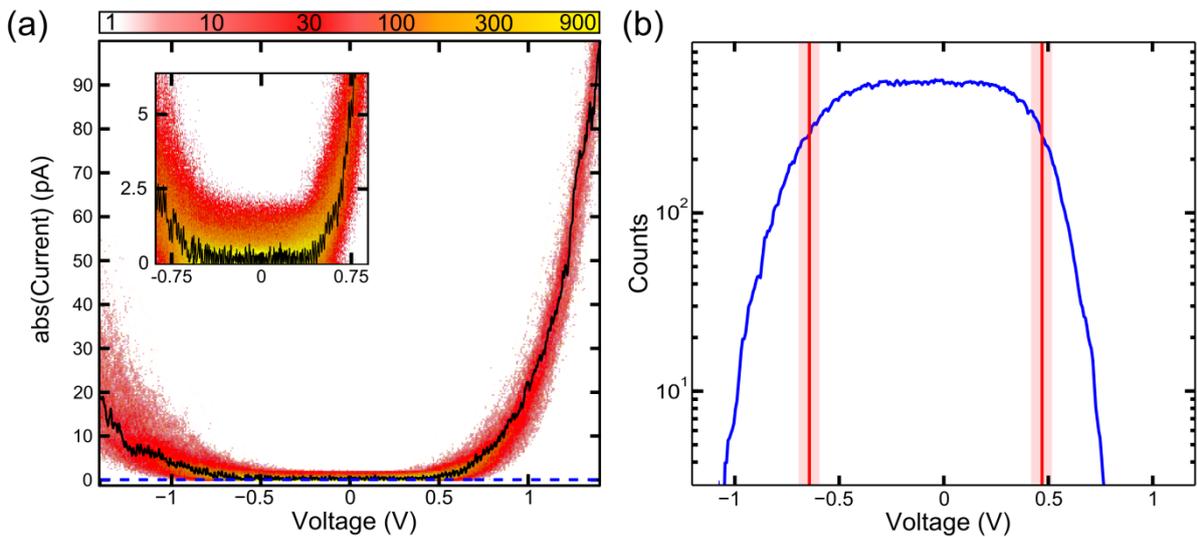

**Figure 2.** (a) Colormap histogram including 205 STS current-voltage characteristics (current in absolute value). A representative current-voltage curve is plotted in solid black line on top. Inset: zoom around zero current in the colormap histogram. (b) One dimensional histogram extracted from a profile along the dashed line (zero current) in the two dimensional histogram (a). The vertical red lines indicate the valence and conduction band values (-0.64 ± 0.06 eV and 0.47 ± 0.06 eV, respectively). The shaded area represents the uncertainty for each value.

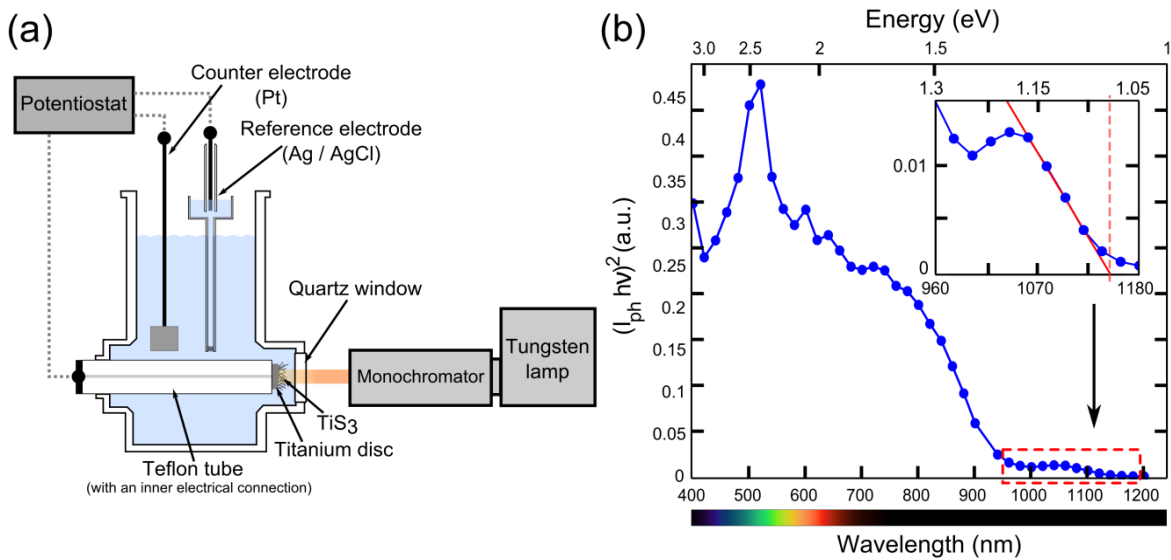

**Figure 3.** (a) Schematic drawing of the experimental setup used for the photoelectrochemical measurements to determine the optical bandgap. (b) Photocurrent density as a function of light wavelength (energy). The optical bandgap energy (1.07 ± 0.01 eV) is determined by the point where the linear fit cuts the zero photocurrent density (highlighted in the inset).

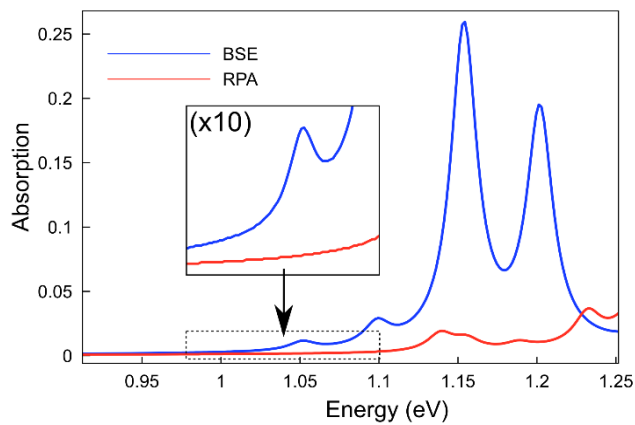

**Figure 4.** Calculated absorption spectra for TiS$_3$ ribbons using RPA (red) and solving the BSE (blue). The first peak in the BSE spectrum indicates the optical bandgap (1.05 eV), while the electronic gap is 1.15 eV and coincides with the first peak of the RPA spectrum. Inset: zoom of the energy region around the absorption peak corresponding to the optical bandgap.

**The electronic and optical bandgap of layered titanium trisulfide ($TiS_3$) are measured with scanning tunnelling spectroscopy and photoelectrochemical experiments.** From these measurements we determine a large exciton binding energy (130 meV) that is 10-100 times larger than that of conventional semiconductors. This value is in good agreement with *ab-initio* calculations.

**Keywords: titanium trisulfide, exciton binding energy, layered semiconductors, scanning tunneling spectroscopy, rpa bse**

Aday J. Molina-Mendoza, Mariam Barawi, Robert Biele, Eduardo Flores, José R. Ares, Carlos Sánchez, Gabino Rubio-Bollinger, Nicolás Agraït, Roberto D'Agosta\*, Isabel J. Ferrer\*, and Andres Castellanos-Gomez\*

**Electronic bandgap and exciton binding energy of layered semiconductor $TiS_3$**

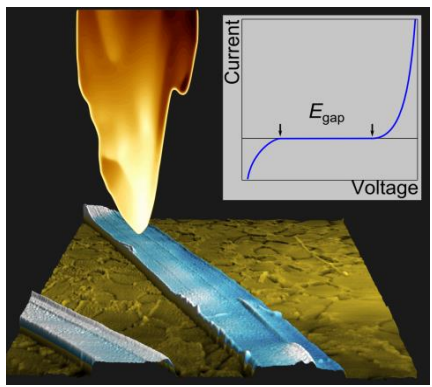

Supporting Information

**Electronic bandgap and exciton binding energy of layered semiconductor TiS$_3$**

*Aday J. Molina-Mendoza, Mariam Barawi, Robert Biele, Eduardo Flores, José R. Ares, Carlos Sánchez, Gabino Rubio-Bollinger, Nicolás Agraït, Roberto D'Agosta\*, Isabel J. Ferrer\*, and Andres Castellanos-Gomez\**

**TiS$_3$ ribbons characterization by AFM**

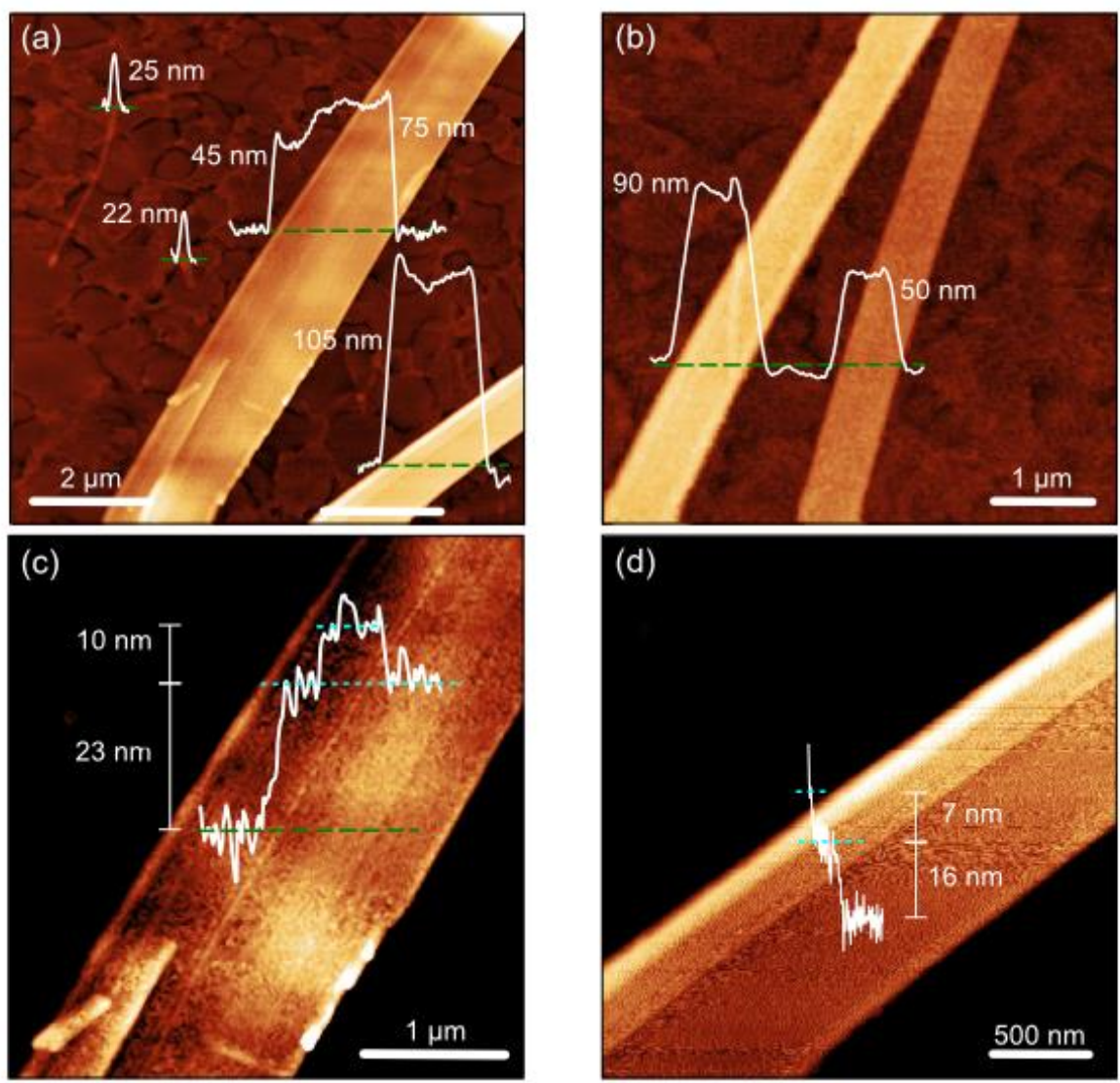

**Figure S1. (a)** and **(b)** AFM topographic images of different TiS$_3$ ribbons on gold substrate with line profile drawn on top. TiS$_3$ ribbons vary in thickness and surface from 25 nm up to 105 nm with average thickness of 60 nm. TiS$_3$-Au interaction is not strong enough to keep the ribbons attached to the substrate during contact mode AFM imaging and therefore the AFM images were performed in non-contact mode to avoid moving the ribbons with the AFM tip. **(c) and (d)** Enhanced contrast AFM topographic images corresponding to part of the TiS$_3$ ribbons in (a) to facilitate the identification of flat terraces and step edges.

**Current-Voltage characteristics dependence on tip-sample distance**

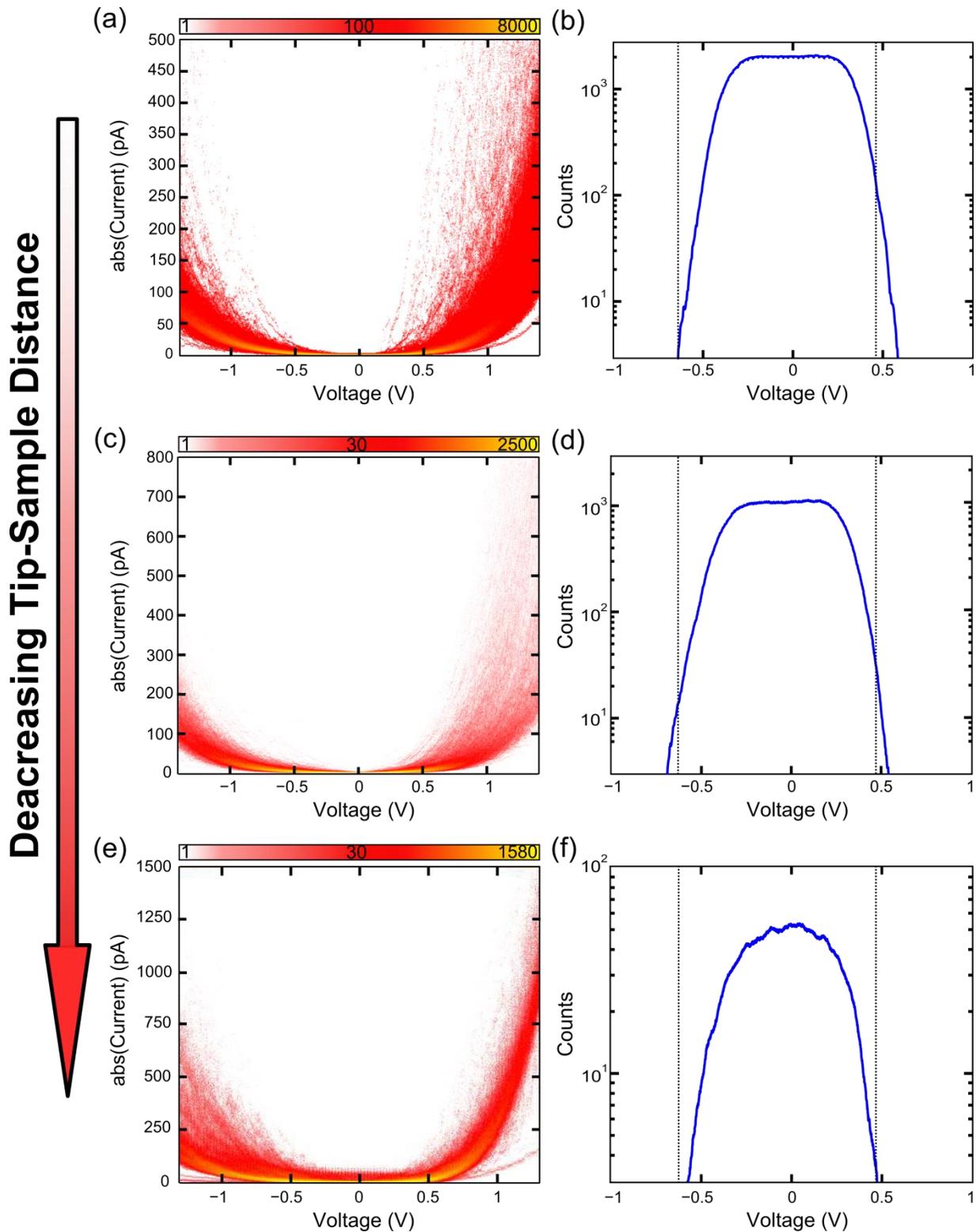

**Figure S2. (a), (c), (e)** Colormap histograms showing STS current-voltage characteristics (current in absolute value) for different tip-sample distances, *i.e.*, different tunneling current setpoints. Tip-sample distance is decreasing from (a) to (e). **(b), (d), (f)** One dimensional histograms extracted from a profile along zero current in the two dimensional histogram on the left of each figure, *i.e*., (b) corresponds to the zero current in (a), (d) corresponds to (c) and (f) corresponds to (e).

## 2D Histograms construction

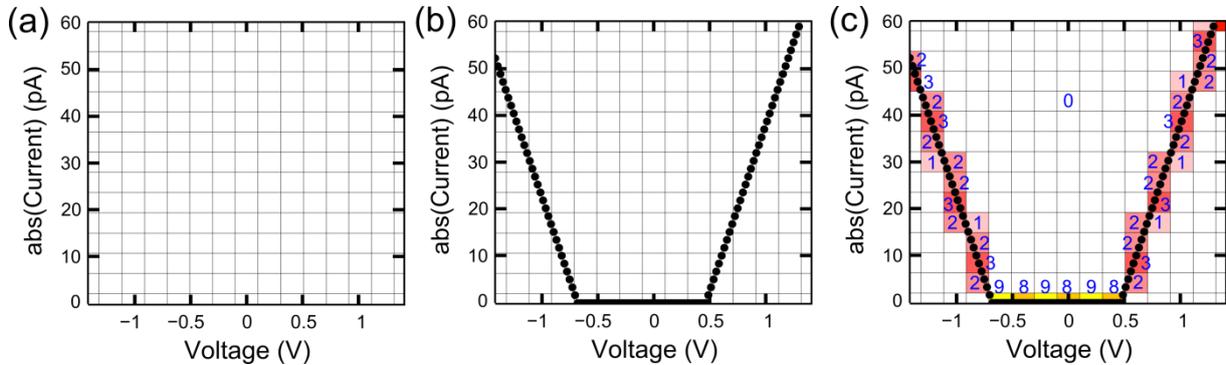

**Figure S3. (a)** Grid representing the bin sizes for a 2D histogram construction. Both current and voltage axes are divided in 15 bins with the same size in order to count the points of a curve that fall within them. **(b)** Synthetic semiconductor-like current-voltage curve in absolute value. **(c)** 2D histogram constructed as explained in the Experimental Section from the artificial current-voltage curve shown in (b). The numbers in blue color indicate the number of datapoints whose (abs(I),V) values are within each bin.

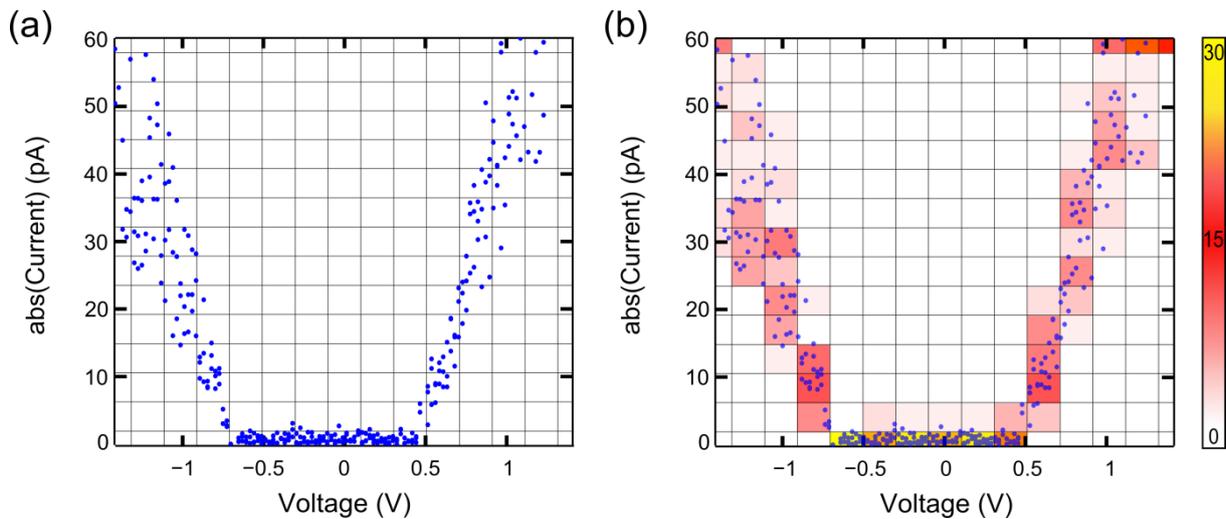

**Figure S4. (a)** Synthetic semiconductor-like current-voltage characteristic with the addition of gaussian white noise. **(b)** 2D histogram (15x15 bins) constructed from three different "noisy" curves as the one shown in (a). The 2D histogram is constructed in the same way as the one in Figure S3.

## TiS$_3$ ribbons atomic structure

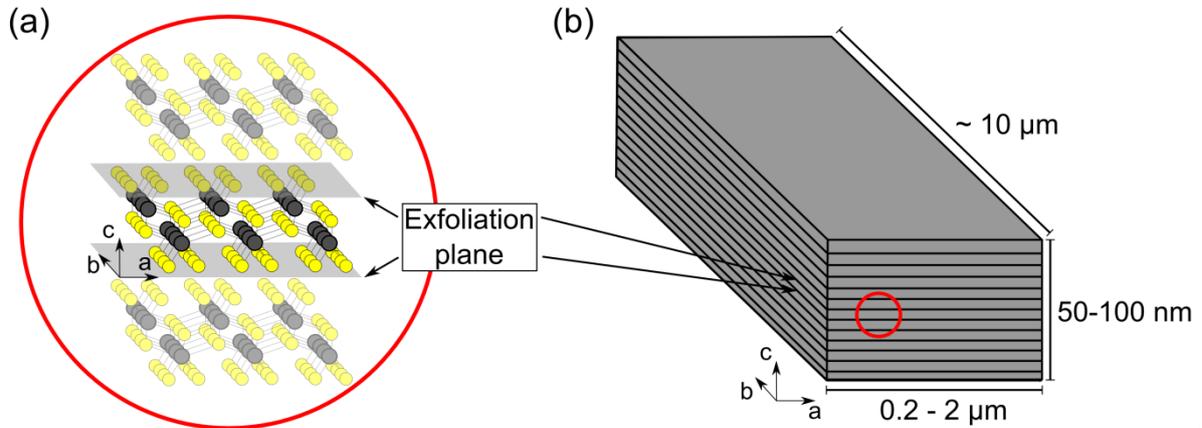

**Figure S5**. **(a) and (b)** Schematic drawing of TiS$_3$ crystal structure. Ti atoms are represented by grey color spheres, while S atoms are represented by yellow spheres. TiS$_3$ is a layered material that grows in parallel sheets with one dimensional chains in the form of stacked triangular prisms. The exfoliation from bulk occurs trough the (001) plane, represented here by grey rectangles in (a) and by solid black lines in (b).

## Differential conductance of TiS$_3$ STS measurements

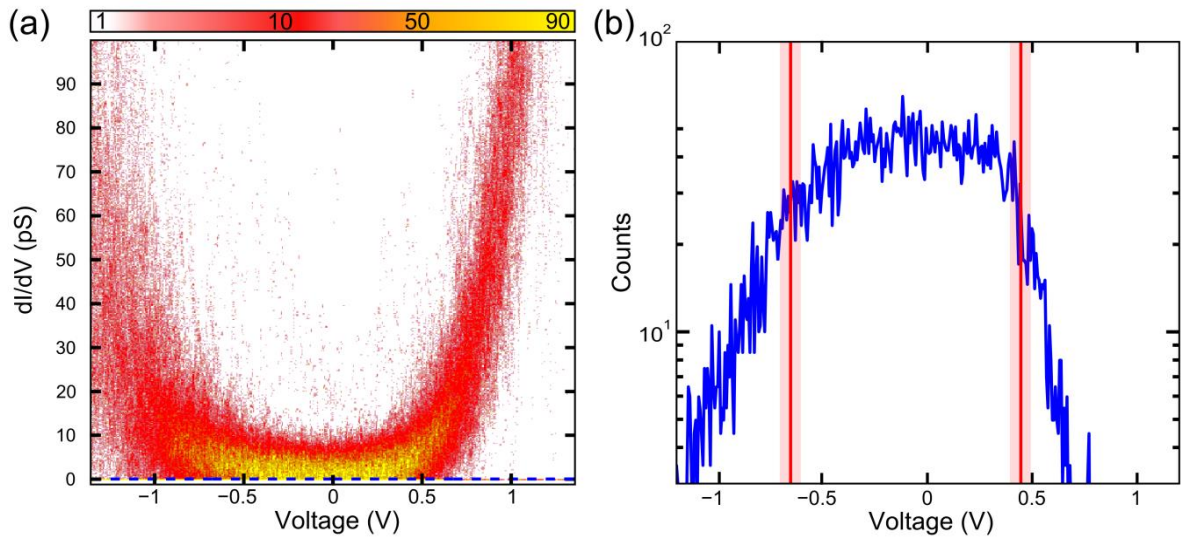

**Figure S6.** **(a)** Colormap histograms showing 205 STS differential conductance-voltage curves calculated from the STS current-voltage characteristics shown in Figure 1a. **(b)** One dimensional histogram extracted from a profile along zero differential conductance (dashed blue line) in the two dimensional histogram shown in (a).